\documentclass[a4paper,amsmath,amssymb,floatfix,prb,footinbib,twocolumn,superscriptaddress,preprintnumbers,showpacs]{revtex4}

\usepackage{amssymb}
\usepackage{graphicx}
\usepackage{amsfonts}
\usepackage{amsmath,bm}
\usepackage{color}
\usepackage{url}
\usepackage[colorlinks=true ,citecolor=blue]{hyperref}

\newcommand{\comment}[1]{}

\begin{document}
\title{Rashba spin-orbit-interaction-based quantum pump in graphene}

\author{Dario Bercioux}
\affiliation{Freiburg Institute for Advanced Studies, Albert-Ludwigs-Universit\"at, D-79104 Freiburg, Germany}
\author{Daniel F. Urban}
\affiliation{Physikalisches Institut, Albert-Ludwigs-Universit\"at, D-79104 Freiburg, Germany}
\author{Francesco Romeo}
\author{Roberta Citro}
\affiliation{Dipartimento di Fisica ``E. R. Caianiello" and Spin-CNR, Universit$\grave{a}$ degli Studi di Salerno, I-84084 Fisciano (Sa), Italy}

\date{\today}

\begin{abstract}
We present a proposal for an adiabatic quantum pump based on a graphene monolayer patterned by electrostatic gates and operated in the low-energy  Dirac regime.
The setup under investigation works in the presence of inhomogeneous spin-orbit interactions of intrinsic- and Rashba-type and allows to generate spin polarized coherent current. A local spin polarized current is induced by the pumping mechanism assisted by the spin-double refraction phenomenon.
\end{abstract}
\maketitle

Graphene-based devices have recently been the object of intense experimental and theoretical studies.  Due to its highly integrable ultra-flat geometry,\cite{Meric:2011} high electron mobility,  tunable carrier concentration,\cite{Das_Sarma:2011} high chemical homogeneity, and excellent intrinsic transport properties,\cite{Novoselov:2004} graphene monolayers are becoming the key technological ingredient in molecular electronics. On the other hand, quasi-particles in the honeycomb lattice of graphene | close to the charge neutrality point | can be described as massless Dirac-Weyl fermions. All these exciting properties make graphene a promising  material for realizing devices having exotic functionalities.
An important class of these devices are the \textit{pure spin-current generators} (PSGs),  being them
conceptual relevant devices in semiconductor spintronics.\cite{Zutic:2004}
Among different spintronics mechanisms that are useful for the design of a PSG, an interesting option is
the implementation of the adiabatic quantum pumping (AQP) technique.\cite{Watson:2003}
However, implementation of PSG schemes is strongly material-dependent and therefore requires
an accurate control of interface properties. Motivated by this requirement,
the application of the AQP to graphene--based devices has become a relevant research activity
in spintronics-oriented molecular electronics.\cite{spin_polarized_qp_graphene}
In addition,  the important role of the spin-orbit interaction (SOI) and its tunability via all-electrical means in graphene-based systems has recently been reported.\cite{varykhalov:2008} Despite the relevance of the SOI for spintronics, the latter has not yet been intensely studied in combination with the AQP technique in order to provide a mechanism to generate  a pure spin current.
%
%
\begin{figure}[!ht]
	\begin{center}
	\includegraphics[width=0.7\columnwidth]{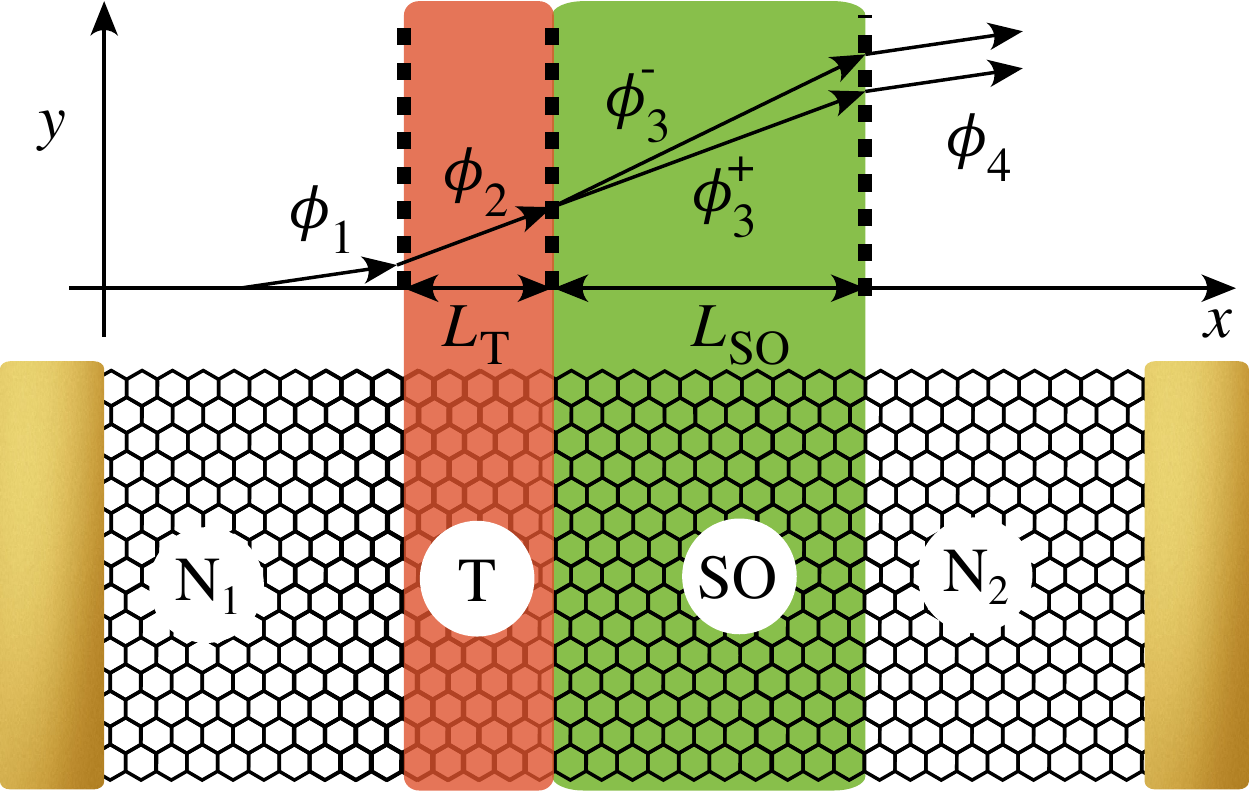}
	\caption{Graphene-based pumping device. The monolayer graphene is sandwiched between two metallic contacts.
An electrostatic top gate (region T) of length $L_\text{T}$ and a back gate (region SO) of length $L_\text{SO}$ are placed between the two contacts, which leave two `normal' graphene regions N$_1$ and N$_2$. Finite spin-orbit interactions are present in region SO, only. The top-part of the figure schematically shows the relevant kinematic angles $\phi_i$ for the scattering process. In the SO region, spin double refraction is observed, \emph{i.e.}, the two spin species move in different directions $\phi_3^\pm$.\label{figure:one}}
	\end{center}
\end{figure}
%
%

Here, we present a proposal of AQP for graphene in the presence of gate tunable SOI | we show this setup may lead to spin-polarized ballistic transport.

Our proposal is based on the exploitation of the spin-double refraction phenomenon\cite{Bercioux:2010} (SDRP), which occurs at the interface between the SOI-active and the normal-graphene regions, in the AQP framework. Other proposal of quantum pumping in graphene recently appeared\cite{prada_2011,tiwari_2010,zhu_2009} but none of these considered the relevant effect of controllable SOIs.
Symmetries in graphene allow for two types of SOIs:\cite{kane:2005} the intrinsic and the Rashba one. The former, originating
from carbon intra-atomic SOI, has been estimated to be rather weak in
clean flat graphene.\cite{Huertas-Hernando:2006} However, a recent theoretical proposal suggests how atoms-mediated hopping can enhance it.\cite{weeks:2011} The Rashba SOI originates
from interactions with the substrate,  
 Stark and/or curvature effects.\cite{Huertas-Hernando:2006} It has been experimentally enhanced  in graphene samples on Ni with intercalated Au atoms.\cite{varykhalov:2008}

We consider a graphene-based system in the low-energy regime assuming that the two Dirac cones can be treated independently, meaning that they are not mixed by interaction effects or boundary conditions.\cite{reviews} Within this approximation the single-valley Hamiltonian for the system reads:
%
%
\begin{align}\label{eq1}
\mathcal{H} = & v_\text{F} (s_0\otimes \vec{\sigma}) \cdot \vec{p} 
+\frac{\lambda(x)}{2}(s_y \otimes \sigma_x-s_x \otimes \sigma_y)\nonumber\\
&+\Delta(x) (s_z \otimes \sigma_z)+V_\text{T}(x)+V_\text{SO}(x).
\end{align}
%
%
Here, the $\sigma_{x/y/z}$  and $s_{x/y/z}$ are Pauli matrices describing the pseudo-spin  and fermionic spin degree-of-freedom, respectively. The matrix $\sigma_0=s_0=\mathbb{I}_2$ is the identity matrix of dimension 2.  In the Hamiltonian \eqref{eq1} the first term describes particles with Dirac-like energy dispersion, the second one is the  Rashba SOI, while the third term takes into account the intrinsic SOI of  graphene.
Let us note that at the lowest order in the long-wavelength approximation, the velocity operator $\vec{v}=v_\text{F}(s_0\otimes \vec{\sigma})$ does not involve spatial derivatives.\cite{Bercioux:2010,Rakyta:2011} The fourth term in~\eqref{eq1} describes an electrostatic gate of length $L_\text{T}$ and strength $V_\text{T}$ placed before the region of finite SOI. The latter term  describes a region of length $L_\text{SO}$ characterized by the presence of a back-gate $V_\text{SO}$ having the two-fold action of (i) varying the electrochemical potential by a quantity $V_\text{SO}$, and (ii) tuning the intensity of the Rashba SOI according to $\lambda(x)\approx\lambda_0(x)+\delta g V_\text{SO}(x)$. Here $\delta g$ is a proportionality constant between the Rashba SOI and the applied electrostatic potential $V_\text{SO}$, while $\lambda_0$ is the bare Rashba SOI.\cite{Huertas-Hernando:2006}

In order to study AQP in this set-up, we allow the electrostatic potentials in~\eqref{eq1} to be time-dependent quantities.
In particular, their time-dependence is chosen as $V_\beta(x,t)=V^0_\beta(x)+\delta V_\beta\sin(\omega t+\varphi_\beta)$ with $\beta \in \{\text{T}, \text{SO}\}$, where
$\varphi=\varphi_\text{SO}-\varphi_\text{T}\neq0$ is a phase difference between the two pumping parameters.
An important consequence of the gate modulation is an induced time-dependence also for the Rashba SOI ($\lambda(x)\to\lambda(x,t)$).

Assuming current conservation laws,\cite{Scheid:2007} the  charge current density $\bm{J}_\text{C}$ and the spin current density $\bm{J}_\text{S}$ can be expressed  as $\bm{J}_\text{C}=q v_\text{F} \psi^{\dag}(s_0\otimes \bm{\sigma})\psi$ and $\bm{J}_\text{S}=\frac{\hbar}{2} v_\text{F} \psi^{\dag}(s_z \otimes \bm{\sigma})\psi$, respectively.
In order to find an expression for the pumped charge and spin currents, we evaluate the scattering matrix $\mathcal{S}$ of the whole system. Thus, we require continuity of the particle wave-function at all potential discontinuities of Eq.~\eqref{eq1} along the $x$ axis.
We consider hereafter a purely electronic transport in the leads, while the SOI parameters (both $\lambda$ and $\Delta$) are taken finite only in the scattering region labeled `SO' in Fig.\ \ref{figure:one}.
Charge and spin currents generated in the system can be evaluated within a modified
scattering field approach.\cite{Buttiker92,Brouwer:1998} The currents in lead $\alpha=1,2$ are obtained by the average over incoming angles $\phi_\text{in}$,
%
%
\begin{align}
\langle\mathcal{J}^{\alpha}_{\gamma}\rangle=\int d \phi_{\text{in}}\,\mathcal{J}^{\alpha}_{\gamma}(\phi_{\text{in}})\, ,
\end{align}
%
$\gamma=\text{C/S}$, and the $\phi_{\text{in}}$--dependent current contributions read
%
%
\begin{align}
\mathcal{J}^{\alpha}_{\gamma}(\phi_{\text{in}})=\left(2\mathcal{N}_{\bot}\sin(\varphi)\cos(\phi_\text{in}) \frac{\omega}{2\pi}\right)\times \nonumber \\
\delta V_\text{T} \delta V_\text{SO}\sum_{\beta s s'}\Lambda^{\gamma}_s \mathrm{Im}\left[\frac{\partial\mathcal{S}^{\alpha\beta}_{ss'}}{\partial {V_\text{T}}} \frac{\partial\mathcal{S}^{\alpha\beta \ast}_{ss'}}{\partial {V_\text{SO}}}\right],
\end{align}
%
%
where the scattering matrix is also a function of  $\phi_{\text{in}}$. Here, a factor of two accounts for the valley degeneracy and
$\mathcal{N}_{\bot}=(E_\text{F} W)/(h v_\text{F})$ is the number of transverse channels.
The constant $\Lambda_s^{\gamma}$ takes the value $\Lambda_s^\text{C}=q$ for the charge current and $\Lambda_s^\text{S}=s\hbar/2$ for the spin current.
Note that the pumped currents are maximal for a phase difference $\varphi=\pi/2$ in the time variation of the two pumping parameters $V_\text{T}$ and $V_\text{SO}$.
In the following we will express the currents in units of $\frac{\omega}{\pi} \mathcal{N}_{\bot} \frac{\delta V_\text{T}}{V_\text{T}^0} \frac{\delta V_\text{SO}}{V_\text{SO}^0}$, with a pre-factor $q$ for the charge current and $\hbar/2$ for the spin currents.
The order of magnitude of the pumped charge current is of the order  $1$--$20$~nA assuming
$\omega/(2\pi)=5$~GHz, $\mathcal{N}_{\bot}=100$ and the weak pumping limit $\delta V_\beta/V_\beta^0 \sim 0.1$.
For the Rashba SOI strength we take the maximal value of $\lambda=13$~meV.\cite{varykhalov:2008}

We start by analyzing the refraction processes at each of the three interfaces N$_1$--T, T--SO, and SO--N$_2$ (c.f.~Fig.~\ref{figure:one}).
Translational invariance along the $y$ direction implies conservation of the momentum along the interfaces, \textit{i.e.}, $k_y(x_\text{interface}^-)=k_y(x_\text{interface}^+)$.
By using simple kinematic considerations~\cite{Bercioux:2010} we can write a general relation expressing the refraction angle at the interfaces. This reads
%
%
\begin{equation}\label{angle:one}
\phi_{i+1}=\arcsin\left[ \frac{k_i}{k_{i+1}}\sin\phi_i\right],
\end{equation}
%
%
where the $k_{i}$ and $\phi_{i}$  (with $i=1,\ldots,4$ and $\phi_1\equiv\phi_\text{in}$) are the momentum modulus and propagation directions in the four regions, respectively (c.f.~Fig.~\ref{figure:one}). At each interface there exists an incoming critical angle $\phi_{i}^\text{c}$ over which the incoming wave is totally refracted along the interface, this is given by
%
%
\begin{equation}\label{angle:two}
\phi_{i}^\text{c}=\arcsin\left[ \frac{k_{i+1}}{k_i}\right].
\end{equation}
%
%

When entering a region of finite Rashba SOI, an incoming wave splits into two outgoing waves with different propagation directions which are associated to the two different spin eigenstates within the Rashba SOI region. This refraction along the different spin channels is called SDRP\cite{Bercioux:2010} and is relevant for the spin polarized transport.
When one of the modes reaches the critical angle, it propagates completely parallel to the interface, while becomes evanescent in the perpendicular direction.\cite{Bercioux:2010}

We present results for two cases: the first one with negligible intrinsic SOI and a second one with a finite intrinsic SOI, still smaller than the Rashba SOI.
In general the spin [charge] pumping current is an odd [even] function of the injection angle $\phi_{\text{in}}$, as can be seen from the symmetry properties of the SOI region.\cite{Bercioux:2010,note:one}
%
%
\begin{figure}[t]
	\begin{center}
	\includegraphics[width=\columnwidth]{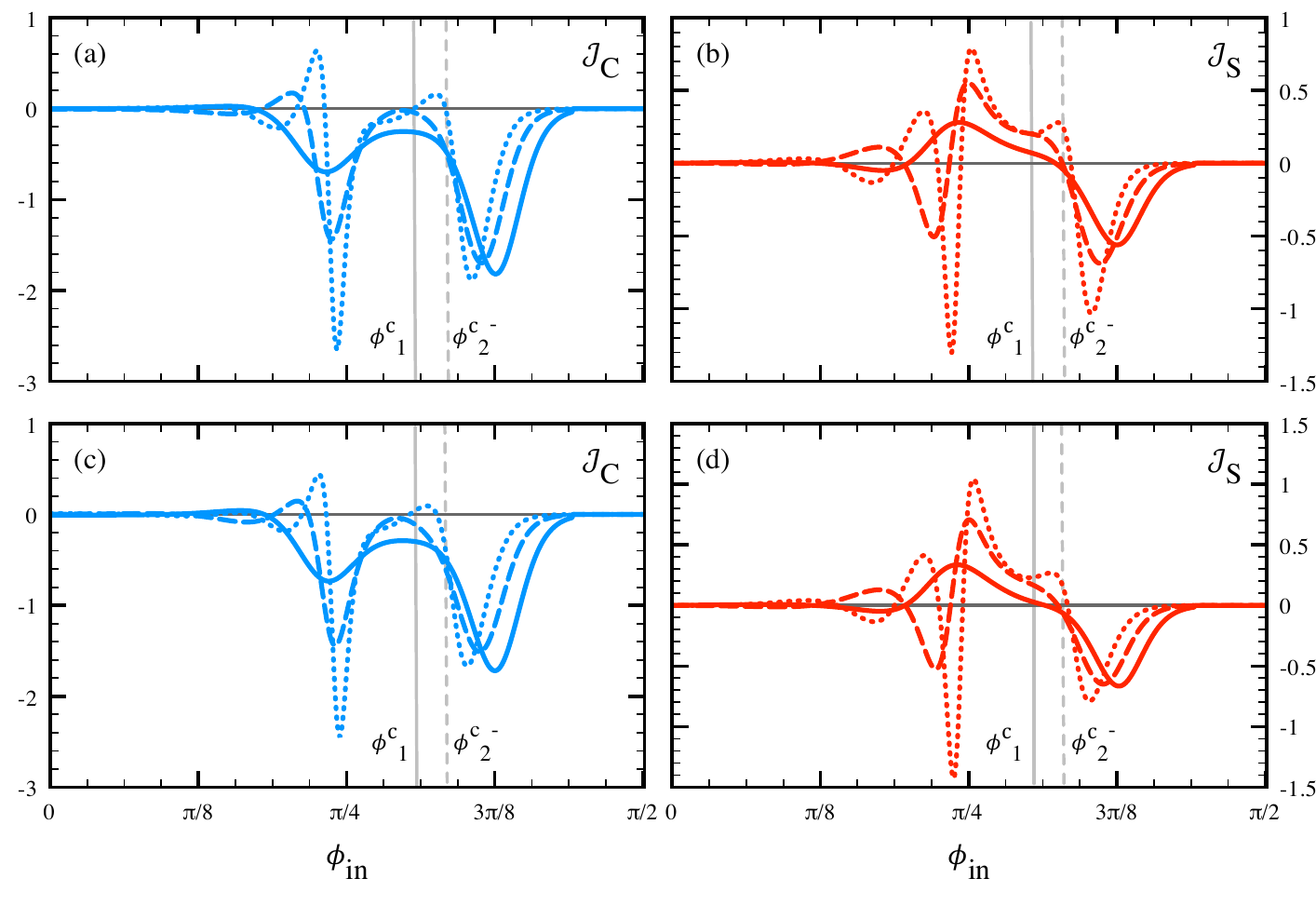}
	\caption{
Charge $\mathcal{J}_\text{C}$ (left panels) and spin $\mathcal{J}_\text{S}$ (right panels) pumping currents as a function of the injection angle $\phi_{\text{in}}\equiv\phi_1$ for different lengths of the SOI region $L_\text{SO}$: (i) 50~nm (solid line), (ii) 75 nm (dashed line), (iii) 100 nm (dotted-dashed line). The others parameters are: $V_\text{T}^0=1.3$~meV, $L_\text{T}=50$~nm, $E_\text{in}=71.5$~meV, $\lambda_0=11.7$~meV,
$\Delta\sim 0$~meV (upper panels) and $\Delta\sim2.6$~meV (lower panels), $V_\text{SO}^0=13$~meV, and $\delta g=0.1$.\label{fig:fig-five}
}
	\end{center}
\end{figure}
%
%

In Fig.~\ref{fig:fig-five} we report the angular dependence of the charge and spin pumping currents for different lengths of the SOI region. For small values of the injection angle, both currents are rather small, then for $\phi_{\text{in}}\gtrsim 0.5$ they both acquire finite values and oscillate as a function of the injection angle. For injection angles greater than the first critical angle of the N$_1$--T interface they become negative with a peak at the smaller of the two critical angles $\phi_2^{c,\pm}$ of the T--SOI interface. For even larger values of $\phi_{\text{in}}$,  the two currents both approach zero. A finite value of the intrinsic SOI increases the overall value of the currents.

\begin{figure}[!b]
	\begin{center}
	\includegraphics[width=\columnwidth]{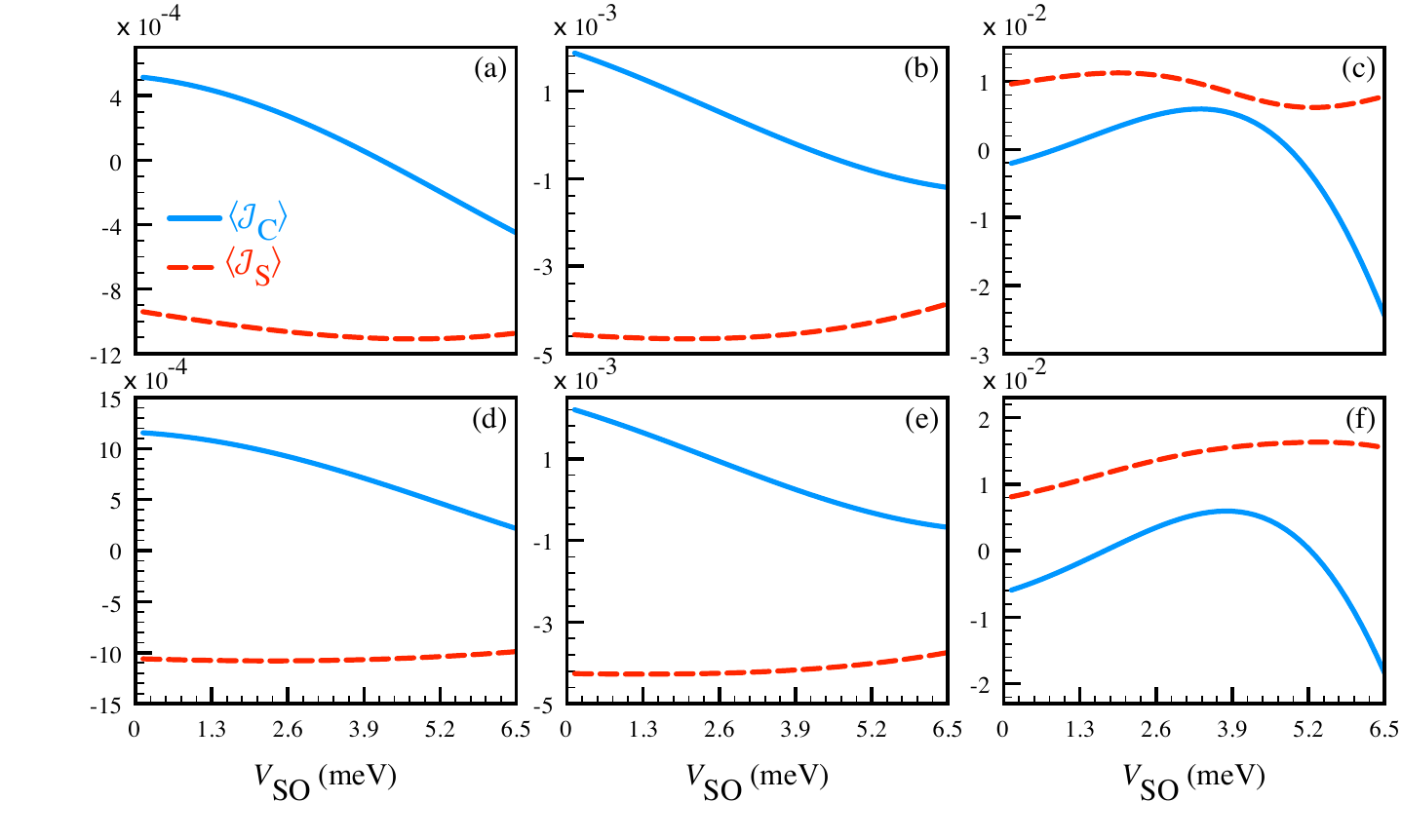}
	\caption{ Angle averaged charge $\langle \mathcal{J}_\text{C}\rangle$ (blue-solid line) and spin  $\langle \mathcal{J}_\text{S}\rangle$ (red-dashed line) pumping currents as a function of the electrostatic gate voltage $V_\text{SO}$ for different lengths $L_\text{SO}$ of the SOI region: (a) and (d) 50~nm, (b) and (e) 75~nm, and (c) and (f) 100~nm. The integration region is $\phi_\text{in}\in[0.4,0.57]$ for (a) and (d), $\phi_\text{in}\in[0.61,0.69]$ for (b) and (e) and $\phi_\text{in}\in[0.9,1.05]$ for (c) and (f). For the upper panels (a)--(c) $\Delta=0$, for the lower ones (d)--(f) $\Delta=2.6$~meV. The other parameters are common to all the panels: $V_\text{T}^0=1.2$~meV, $L_\text{T}=50$~nm, $E_\text{in}=71.5$~meV,  $\lambda_0=11.7$~meV, and $\delta g=0.1$.\label{fig:fig-seven}
}
	\end{center}
\end{figure}

The regions where the charge current changes sign are particularly interesting. There, by performing an integration of both currents over a small range of injection angles, it is possible to find a considerable  value for the pumped spin current while the pumped charge current is negligible. 
The evaluation of the average currents by integration over a small range of angles can be experimentally implemented by using a quantum point contact (QPC) far away from the scattering region. By changing  the width of the QPC, which changes the angular structure of the electron flow, it is possible to change the angular contribution to the transmission through the contact itself.\cite{Topinka:2000}
The data shown in Fig.~\ref{fig:fig-five}  allows to identify distinct regions with this property to be considered as working points of a PSG (c.f.\ caption of Fig.~\ref{fig:fig-seven} for their actual value).
%

%
%
%
The average pumping currents are shown in Fig.~\ref{fig:fig-seven} as a function of the electrostatic voltage $V_\text{SO}$ for different lengths of the SOI region with and without intrinsic SOI.
We recall the twofold action of $V_\text{SO}$ which produces a chemical potential shift accompanied by a variation of the Rashba SOI strength. The main result is an increase of the overall currents for an increase of the SOI region length. In addition, we observe the existence of optimal working points in terms of $V_\text{SO}$ for the PSG in which the average charge pumping current is zero while the spin current is finite. For the case of a 100~nm SOI region [panels (c) and (f)] we can identify two of these working points. Furthermore, we observe that a finite value of the intrinsic SOI is producing only a change in the magnitude of the currents.

In conclusion, we have studied a graphene-based spintronic pump exploiting the SDRP. This is due to the inhomogeneous Rashba SOI that can be obtained via an appropriate spatial modulation of the Au atoms between the graphene layer and the Ni substrate.\cite{varykhalov:2008} Special signatures of the critical refraction angles have been identified in the angle-resolved  properties of the pump. Charge and spin current curves as a function of the gate voltages support the possibility to obtain local spin-polarized transport in the absence of a charge current. An imaging technique similar to the one described by Topinka \emph{et al.}\cite{Topinka:2000} can be used to detect the coherent spin-polarized particle flow with angular resolution. 


We acknowledge N. Andrei, A. Di Bartolomeo and L. Lenz for useful discussions. The work of DB is supported  by the Excellence Initiative of the German Federal and State Governments.

\bibliographystyle{prsty}


\end{document}